\documentclass[twocolumn,showpacs,preprintnumbers,amsmath,amssymb]{revtex4}
\usepackage{amsmath,amsfonts,latexsym,amssymb,graphics,epsfig,subfigure,color,makeidx}
\usepackage{xcolor}
\usepackage[utf8]{inputenc}
\usepackage{graphicx,hyperref}
\usepackage{bm}
\usepackage{color}
\usepackage{multirow}

\begin{document}
	
	\title{Fate of initially bound timelike geodesics in spherical boson stars}
	\author{Yu-Peng Zhang$^{1,2}$\footnote{zyp@lzu.edu.cn},
		Shi-Xian Sun$^{1,2}$\footnote{sunshx20@lzu.edu.cn},
		Yong-Qiang Wang$^{1,2}$\footnote{yqwang@lzu.edu.cn},
		Shao-Wen Wei$^{1,2}$\footnote{weishw@lzu.edu.cn},
        Pablo Laguna$^{3}$\footnote{pablo.laguna@austin.utexas.edu},
		Yu-Xiao Liu$^{1,2}$\footnote{liuyx@lzu.edu.cn, corresponding author}
	}
	\affiliation{
        $^{1}$Lanzhou Center for Theoretical Physics, Key Laboratory of Quantum Theory and Applications of MoE, Key Laboratory of Theoretical Physics of Gansu Province, Lanzhou University, Lanzhou 730000, China\\
        $^{2}$Institute of Theoretical Physics \& Research Center of Gravitation, Lanzhou University, Lanzhou 730000, China\\
        $^3$Center of Gravitational Physics, Department of Physics, University of Texas at Austin, Austin, TX 78712, U.S.A.
	}
	
	\begin{abstract}
		
	Boson stars are horizonless compact objects and they could possess novel geodesic orbits under the equilibrium assumption, which differ from those in black hole backgrounds. However, unstable boson stars may collapse into black holes or migrate to stable states, resulting in an inability to maintain the initially bound geodesic orbits within the backgrounds of unstable boson stars. To elucidate the fate of initially bound geodesic orbits in boson stars, we present a study of geodesics within the spherical space-times of stable, collapsing, and migrating boson stars. We focus on timelike geodesics that are initially circular or reciprocating. We verify that orbits initially bound within a stable boson star persist in their bound states. For a collapsing boson star, we show that orbits initially bound and reciprocating finally either become unbound or plunge into the newly formed black hole, depending on their initial maximal radii. {For initially circular geodesics, we have discovered the existence of a critical radius. Orbits with radii below this critical value are found to plunge into the newly formed black hole, whereas those with radii larger than the critical radius continue to orbit around the vicinity of the newly formed black hole, exhibiting nonzero eccentricities}. For the migrating case, a black hole does not form. In this case, the reciprocating orbits span a wider radial range. For initially circular geodesics, orbits with small radii become unbound, and orbits with large radii remain bound with nonvanishing eccentricities. This geodesic study provides a novel approach to investigating the gravitational collapse and migration of boson stars.
		
	\end{abstract}
	\maketitle
	\section{Introduction}

Since the first detection of gravitational waves from the coalescence of a binary black hole~\cite{Abbott2016a}, more than a hundred new gravitational-wave events have been observed by LIGO and VIRGO~\cite{LIGOScientific:2018jsj,LIGOScientific:2020kqk,LIGOScientific:2021psn}. These observations, together with the shadow observations of the black holes in M87*\cite{eth2019} and Sagittarius $A^*$\cite{eth2022}, are giving us an outstanding opportunity to gain a complete understanding of the nature of black holes, including the dynamics of their horizons.

The evidence from gravitational-wave detections strongly supports black holes and neutron stars as their sources and Einstein's theory of general relativity as the correct theory of gravity. There is, however, still room for considering alternative theories of gravity and exotic compact objects that mimic black holes. Boson stars are the well-known mimickers of black holes and initially proposed in 1960s ~\cite{Feinblum1968,Kaup1968,Ruffini1969}. They have been implicated as potential candidates ~\cite{Bustillo2021} of the GW190521 event~\cite{LIGOScientific:2020iuh}. Since the boson stars were proposed, unraveling the nature of boson stars relies on figuring out their formation mechanisms and assessing their dynamical stabilities \cite{Seidel1994,Cunha:2017qtt,Giovanni2018,Sanchis2019,Cardoso:2019rvt,Choptuik:2019zji,Sanchis-Gual:2021edp,Siemonsen:2020hcg,Sanchis-Gual:2021phr}. Nonlinear numerical evolutions of boson stars~\cite{Sanchis2019,Choptuik:2019zji,Sanchis-Gual:2021edp,Siemonsen:2020hcg,Sanchis-Gual:2021phr} have revealed three distinct outcomes: 1) unstable boson stars collapsing into black holes, 2) unstable boson stars migrating to the stable configurations, and 3) stable boson stars.

{It has been demonstrated that geodesics serve as a powerful tool for studying both static and fully dynamical space-times by exploiting the corresponding incompleteness of geodesic trajectories \cite{Penrose1965,Wald1984}. Null geodesics also play a crucial role in identifying unexpected characteristics in the formation of a common apparent horizon during the head-on collision of binary black holes \cite{1995Sci...270..941M}. They can additionally provide insight into the gravitational collapse of neutron stars \cite{Vincent:2012kn} and the collision of binary black holes \cite{Bohn:2014xxa} through numerical relativity simulations. While neglecting the stability of boson stars, geodesics have been utilized to explore the properties of boson stars under the assumption of equilibrium. The horizonless nature of boson stars ensures the existence of novel orbits \cite{philippe2014,Grould12017,Collodel2018,Yuzhang2021,Herdeiro2021lwl}.
	
Focusing on the stability of boson stars, Cunha et al. \cite{Cunha:2017qtt} have established a close correlation between the stable light-ring and the stabilities of rotating boson stars. Using nonlinear numerical evolutions of ultracompact rotating boson stars, Ref. \cite{Cunha:2022gde} further explored the fate of the light-ring instability of rotating boson stars and presented two potential outcomes: migrating to nonultracompact configurations or collapsing into black holes, resulting in the stable light-ring ultimately disappearing. Therefore, the dynamical changes in space-time will inevitably affect the properties of geodesic orbits within its background. Moreover, when unstable boson stars collapse into black holes or migrate to stable states, the state of initially bound geodesic orbits in the unstable backgrounds cannot be maintained.
	
Note that the analysis regarding the existence of the light-ring in terms of the effective potential under the adiabatic approximation \cite{Cunha:2022gde} cannot fully elucidate the process of how an initially stable light-ring disappears and whether it escapes or plunges into the newly formed black hole. To determine the possible fate of the initially bound geodesic orbits in boson stars, it is necessary to conduct nonlinear simulations of boson stars and investigate the dynamical behaviors of the initially bound geodesics simultaneously. In this paper, we will undertake fully nonlinear evolutions of boson stars and compute the geodesics in the resulting dynamical space-time using the $3+1$ formalism \cite{Vincent:2012kn,Bohn:2014xxa}. The dynamical geodesics will enable us to map the space-time of the boson star during its gravitational collapse and migration. For simplicity, we will focus on spherical boson stars and initially bound time-like geodesics. }

	\section{Methodology} \label{sec:fundamentals}
	
The action used for the boson star is \cite{Sanchis-Gual:2021phr}
	\begin{equation}
	S = \int{d^4 x \sqrt{-g}
		\left(\frac{R}{16\pi} - \frac{1}{2}\partial_\mu\Phi\partial^\mu\Phi^* -V(|\Phi|^2)\right)},
	\label{action}
	\end{equation}
where $\Phi$ is a complex scalar and the scalar potential is given by $V=\frac{1}{2}\mu^2|\Phi|^2+\frac{\lambda}{4} |\Phi|^4$. The units are such that $G=c=1$, and the mass parameter of the scalar field is set to $\mu = 1$. With an appropriate choice of the coupling parameter $\lambda$, one can construct initial configurations of boson stars that are stable, unstable collapsing into a black hole, and unstable migrating into a stable boson star~\cite{Sanchis-Gual:2021phr}.

We assume that the boson star is initially stationary, thus~\cite{Sanchis-Gual:2021phr},
	\begin{equation}
	\Phi=\phi(r)\exp\left(i\omega t\right)
	\end{equation}
and the metric is given by \cite{Sanchis-Gual:2021phr}
	\begin{equation}
	ds^2=-e^{2f_0(r)}dt^2+e^{2f_1(r)}\left(dr^2+r^2\,d\Omega^2\right)
	\label{metric-sbs}
	\end{equation}
with $d\Omega^2 = d\theta^2+\sin^2\theta d\varphi^2$. The field equations for this gravity system become ordinary differential equations for the radial functions $f_0(r)$ and $f_1(r)$, together with an equation for $\phi(r)$. The only input needed to construct the initial boson star configuration is the parameters $\lambda$ and $\omega$.

\begin{figure}[htbp]
	\begin{center}
	\includegraphics[width=0.8\linewidth]{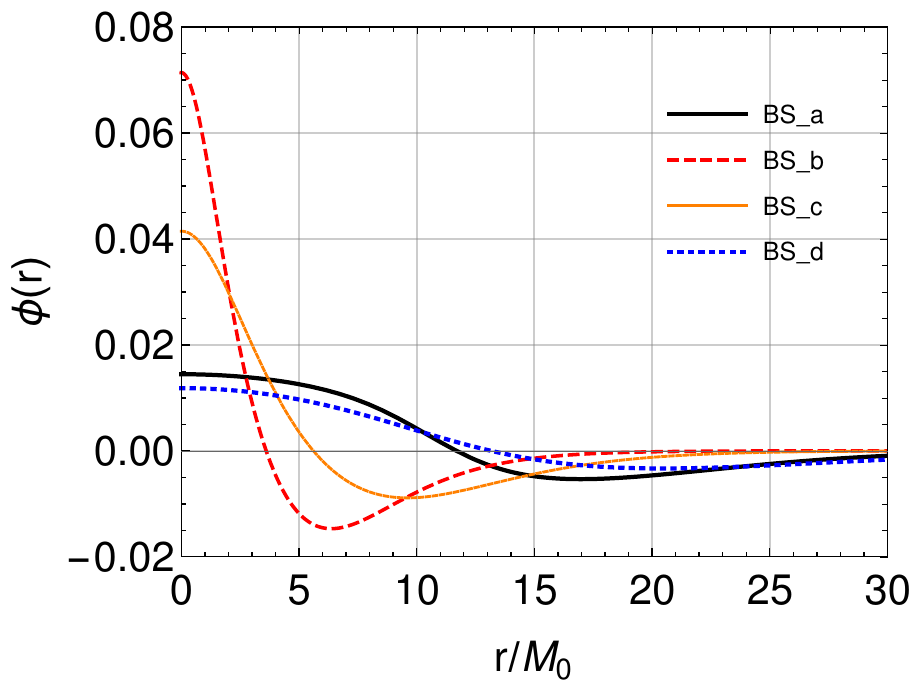} \hspace{5mm}
	\end{center}
	\vspace{-5mm}
	\caption{Initial profiles of the scalar field $\phi(r)$ for the models listed in Table \ref{three_bss}.}
	\label{phi_r_ini}
    \end{figure}

We consider three types of boson stars: i) a stable boson star, ii) an unstable boson star that collapses into a black hole, and iii) an unstable boson star that migrates into another boson star. The parameters $\lambda$ and $\omega$ for each configuration are provided in Table \ref{three_bss}. The Arnowitt-Deser-Misner mass $M$ for each boson star and the mass $M_h$ of the black hole for the collapsing cases are also included in Table \ref{three_bss}. The corresponding profiles for $\phi(r)$ are depicted in Fig. \ref{phi_r_ini}.

	\begin{table}[!htb]
		\begin{center}
			\caption{The coupling parameter $\lambda$, frequency $\omega$, ADM mass $M$, outcome, and mass $M_h$ for the unstable collapsing boson stars. All quantities are given in units of $M_0=\mu^{-1}$.}
			\begin{tabular}{ l c  c  c  c  c }
				\hline
				\hline
				Case&~$\lambda\,M_0^2/(4\pi)$~&~~$\omega\,M_0$~~&~~$M/M_0$~~&~~~~Outcome~~~~&~~$M_\text{h}/M_0$~~ \\
				\hline
				BS\_a      &100     &  0.92          &   2.194         &  stable & --- \\
				BS\_b      &0       &  0.88          &   1.357         &  collapse & 1.246\\
				BS\_c      &0       &  0.92          &   1.284         &   collapse &1.254\\
				BS\_d      &50      &  0.96          &   1.828         &  migration & --- \\
				\hline
				\hline
			\end{tabular}
			\label{three_bss}
		\end{center}
	\end{table}

Given the initial data for each boson star, the space-time evolution is carried out with our spherical numerical relativity code. The code solves the Baumgarte–Shapiro–Shibata–Nakamura formalism in spherical coordinates~\cite{Montero:2012yr} in terms of the following metric
\begin{eqnarray}
ds^2&=&(-\alpha^2+\beta^r\beta_r)dt^2+2\beta_r dt dr\nonumber\\
&+& e^{4\chi}\left(a(t,r)dr^2+r^2 b(t,r)d\Omega^2\right),
\end{eqnarray}
where $\alpha$ is the lapse function and $\beta^i=(\beta^r, 0, 0)$ is the shift vector. For gauge conditions, we use the non-advective $1 + \log$ for the lapse function $\alpha$ \cite{Bona:1997hp}, and a variation of the gamma-driver condition for the shift vector $\beta^r$ \cite{Alcubierre:2002kk}. We impose radiative boundary conditions \cite{Alcubierre:2002kk} to reduce the unphysical reflections at the outer boundary. We adopt the fourth-order Kreiss-Oliger dissipation \cite{Kreiss-Oliger} for stability. Temporal updating is done via method-of-lines with a second-order partially implicit Runge-Kutta method \cite{Montero:2012yr}.

\begin{figure*}[htbp]
	\begin{center}
	\includegraphics[width=0.45\linewidth]{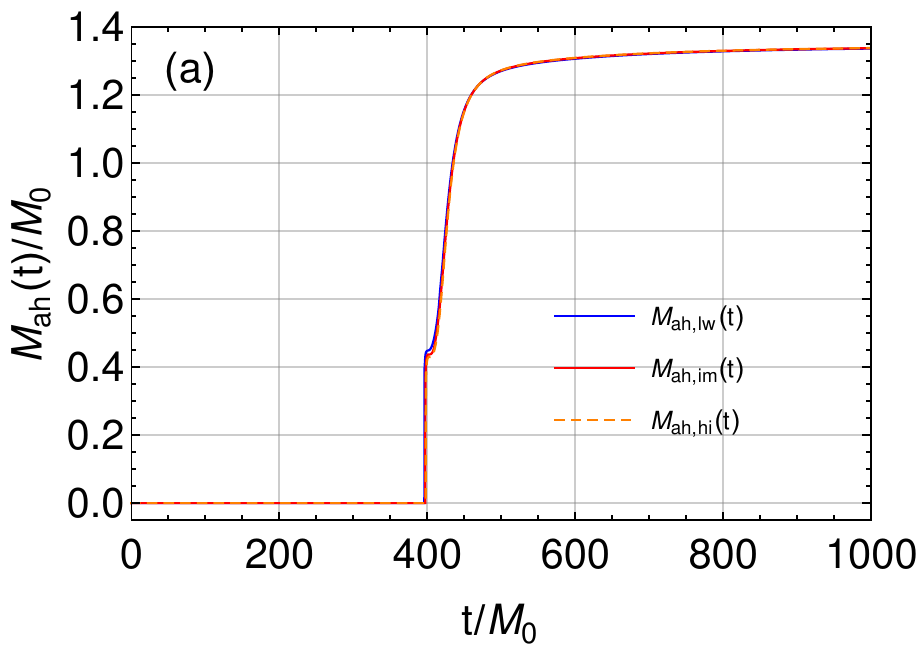}
	\includegraphics[width=0.47\linewidth]{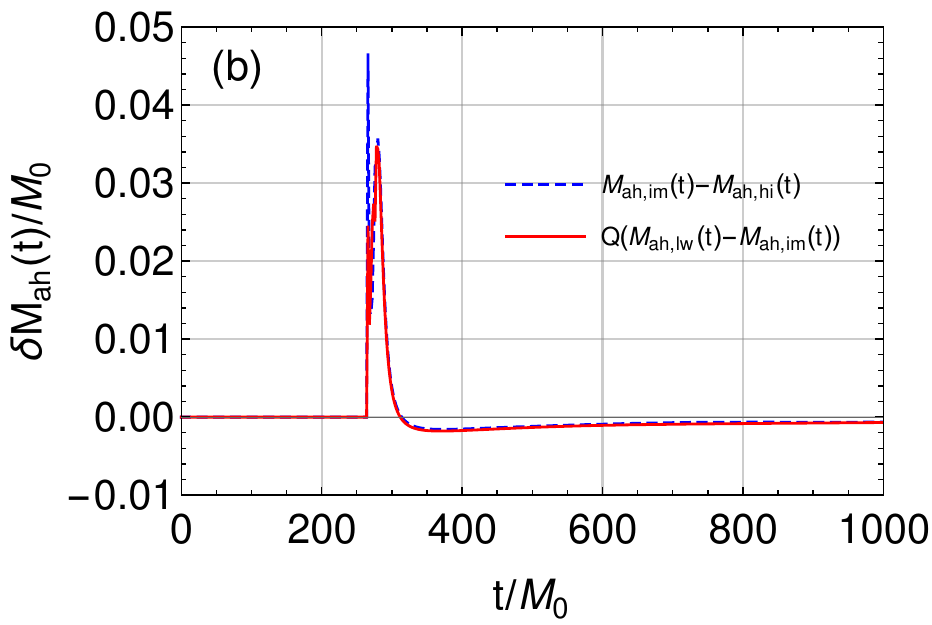}
	\hspace{5mm}
	\end{center}
	\vspace{-5mm}
	\caption{Plots of the convergence test in terms of the irreducible mass of the apparent horizon mass $M_\text{ah}$ for the case BS\_b. The subfigure (a) stands for $M_\text{ah}$ as functions of time with three resolutions $h_{\text{lw}}=0.05M_0$, $h_{\text{im}}=0.04M_0$, $h_{\text{hi}}=0.03M_0$, the subfigure (b) stands for the corresponding differences in the coarse-medium resolution and medium-high resolution. We observe that the factor $Q$ satisfies $Q\simeq 1.65$.}
	\label{convergence}
\end{figure*}

{
Before presenting the results, we briefly discuss the convergence of our code by evolving the unstable boson star case BS\_b with three resolutions: \begin{equation}
h_{\text{lw}}=0.05M_0,~h_{\text{im}}=0.04M_0,~h_{\text{hi}}=0.03M_0.
\label{resolution}
\end{equation}
In principle, the numerical results of a system should converge according to the following rule:
\begin{equation}
\phi_{h} - \phi\propto h^n,\label{convergencerule}
\end{equation}
where $\phi_h$ and $\phi$ represent the numerical results and the exact solution of the system, $h$ corresponds to the resolution, and $n$ denotes the convergence order. Combining Eq. \eqref{convergencerule} with Eq. \ref{resolution}, we have
	\begin{eqnarray} \frac{\phi_{h_{\text{lw}}}-\phi_{h_{\text{im}}}}{\phi_{h_{\text{im}}}-\phi_{h_{\text{hi}}}}=\frac{h_{\text{lw}}^n-h_{\text{im}}^n}{h_{\text{im}}^n-h_{\text{hi}}^n}=Q.
	\label{convergencefactor}
	\end{eqnarray}

We compare the values of the irreducible mass of the apparent horizon $M_\text{ah}$ with different resolutions and provide the corresponding convergence plots in Fig. \ref{convergence}. It is observed that the factor $Q$ approximately equals $1.65$. Substituting resolutions \eqref{resolution} into Eq. \eqref{convergencefactor} and the value of $Q$ yields:
\begin{equation}
Q\simeq 1.65 \simeq \frac{(0.005)^n-(0.004)^n}{(0.004)^n-(0.003)^n}|_{n=3}.
\end{equation}
This implies that the convergence order is about $3$. Although the fourth-order derivative stencils have been employed in our code, the interpolation schemes used for computing the irreducible mass of the apparent horizon impact the corresponding convergences, indicating that the convergence order of our code is approximately $3$.
}

\begin{figure}[htbp]
		\begin{center}
			\includegraphics[width=0.8\linewidth]{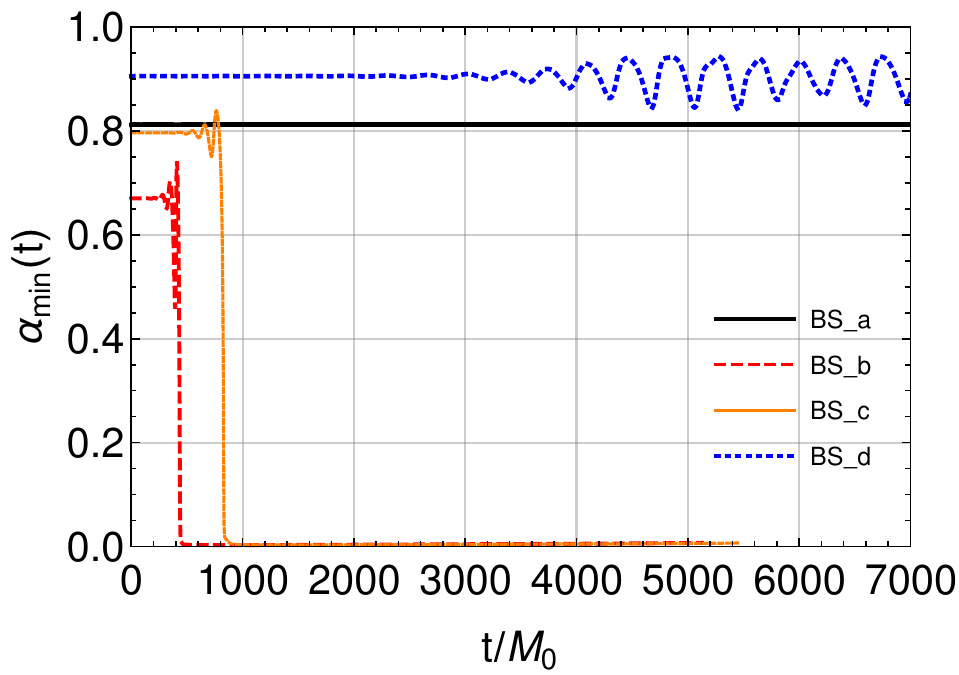} \hspace{5mm}
		\end{center}
		\vspace{-5mm}
		\caption{Evolution of minimum value of the lapse function $\alpha_{\text{min}}$ for the models listed in Table \ref{three_bss}.}
		\label{p_alp}
	\end{figure}

In what follows, we let the computational grid extend into a radius of  $2,000 M_0$ with a grid-spacing $\delta r=0.05 M_0$ and a time-step $\delta t=0.002 M_0$.  Since this work focuses on geodesics, we only show in  Fig. \ref{p_alp} the evolution of the lapse function for each of the models listed in Table \ref{three_bss}. The lapse function remains constant for the stable case, BS\_a. As expected, the lapse function collapses for cases BS\_b and BS\_c, signaling the formation of a black hole. The oscillatory behavior is for the migrating case BS\_d, where the boson star transforms into the stable configuration.

Geodesic integration is calculated \emph{in situ}. That is, the geodesic equations are integrated in parallel with the space-time integration by interpolating the metric from the numerical grid to the geodesic~\cite{Vincent:2012kn,Bohn:2014xxa}. A fourth-order Runge-Kutta temporal integration is used to solve the geodesic equations.
Regarding the initial conditions of the geodesics, for simplicity, we only consider time-like geodesics initially bounded and in the equatorial plane. Two types of initial conditions are investigated: initially circular orbits and radial (reciprocating) orbits. Although the unstable boson star will either collapse into a black hole or migrate to other state, it is initially, to a good approximation, stationary and possesses a timelike killing vector $\xi^\mu=(\partial_t)^\mu$ and a spacelike killing vector $\eta^\mu=(\partial_\varphi)^\mu$. Thus we can set the initial data of the geodesics in terms of the initial energy $E/m = -(\partial_t)^\mu u_{\mu}$, the angular momentum $J/m^2 =(\partial_\varphi)^\mu u_{\mu}$ ($m$ is the mass of the test particle), and the radial effective potentials based on the decomposition of the radial velocity \cite{Grould12017,Collodel2018,Yuzhang2021}. Because we are only considering initially circular and reciprocating orbits, the initial conditions of orbits are fully characterized by their initial radii.


\section{Results} \label{sec:fundamentals}

\begin{figure*}[htbp]
		\begin{center}
			\includegraphics[width=0.45\linewidth]{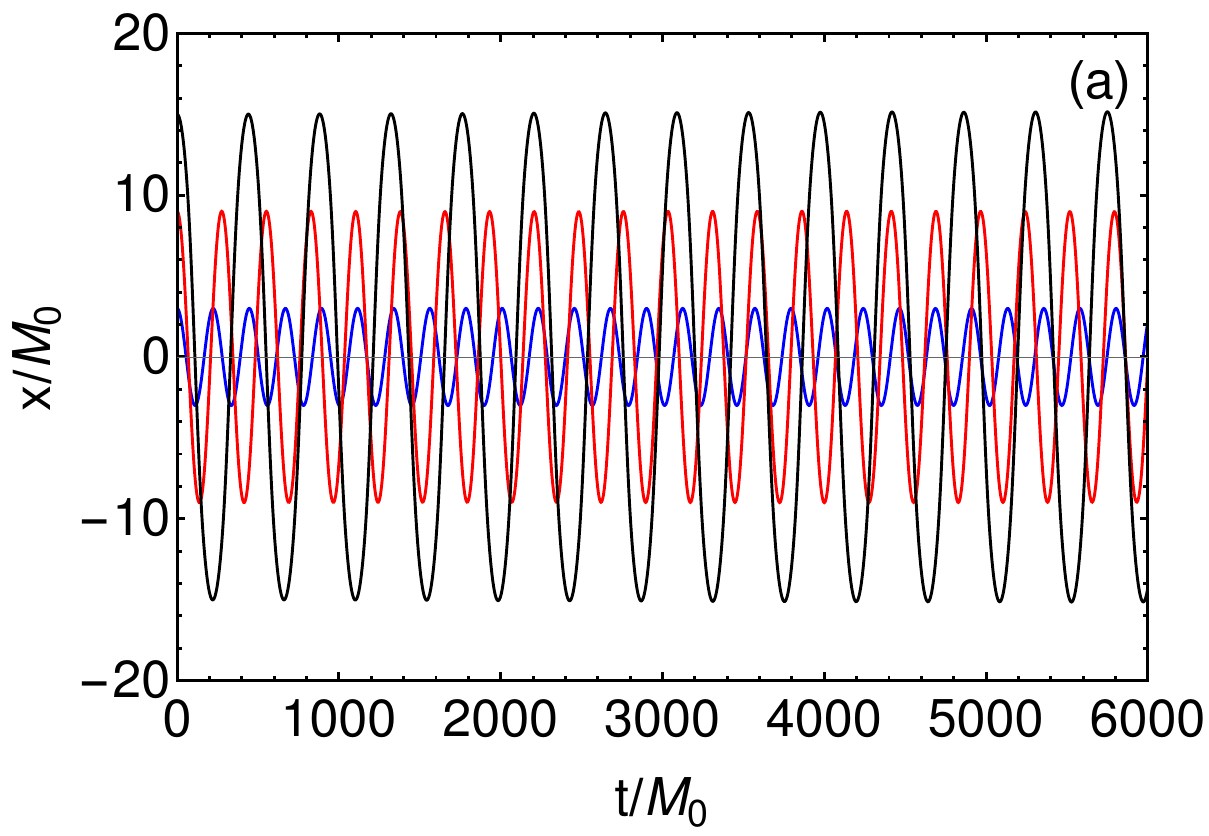}
			\includegraphics[width=0.45\linewidth]{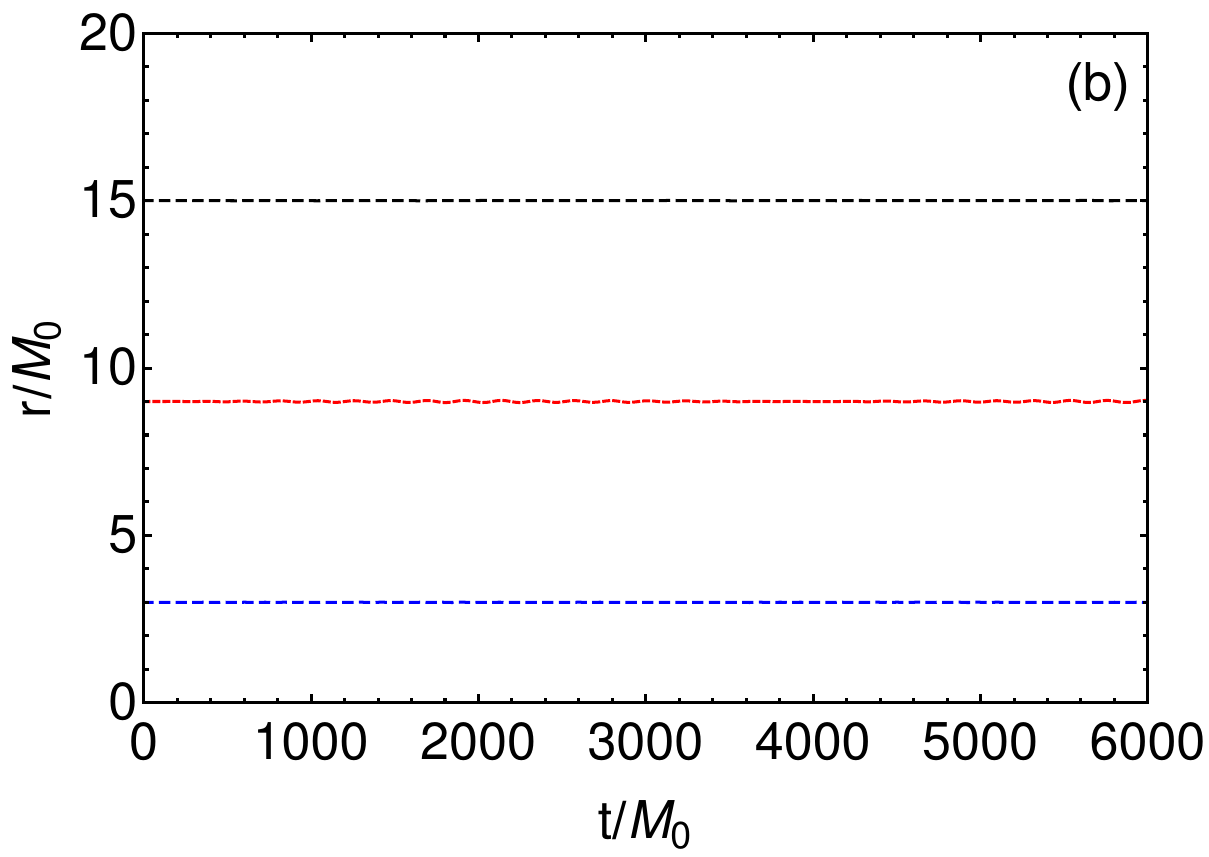} \hspace{5mm}
		\end{center}
		\vspace{-5mm}
		\caption{Evolution of the radial distance for circular and reciprocating geodesics in the space-time of case BS\_a. Panel (a) is for the initially reciprocating orbits, and panel (b) is for the initially circular orbits, respectively. Here, we choose $x=r \cos(\varphi)$.}
		\label{orb_092_l100}
	\end{figure*}

We will first discuss the results for the case of initially bounded geodesics in the stable space-time of BS\_a. We consider geodesics with three different maximal radii $ r/M_0= \lbrace 3, 9, 15 \rbrace$ for both of the reciprocating and circular orbits. Figure~\ref{orb_092_l100} shows the radial distance of the orbit as a function of time. Solid lines are for the reciprocating orbits and dash lines for circular orbits. Since the boson star is stable, i.e., the space-time is not dynamical, the orbits remain reciprocating with the same maximum and minimum radial distance and circular with the constant radius. That is, the initially bounded geodesic orbits remain unchanged.
	
Next, we focus on the unstable boson stars: boson stars collapsing into black holes or migrating into other states. For unstable collapsing boson stars, we consider the cases BS\_b and BS\_c, with the latter has a longer lifetime than the former. As mentioned before, the orbits are initially bounded. There are only three outcomes: i) the orbit becomes unbound from the final black hole; ii) the orbit plunges into the black hole; iii) the orbit remains in a stable path around the black hole. To be able to get a reasonable map of the space-time, we consider a family of geodesics, for both the circular and reciprocating cases, with initial radii in the range $r/M_0\in[3, 21]$ at intervals of $\Delta r/M_0=0.1$.
	
\begin{figure*}[htbp]
		\begin{center}
			\includegraphics[width=0.46\linewidth]{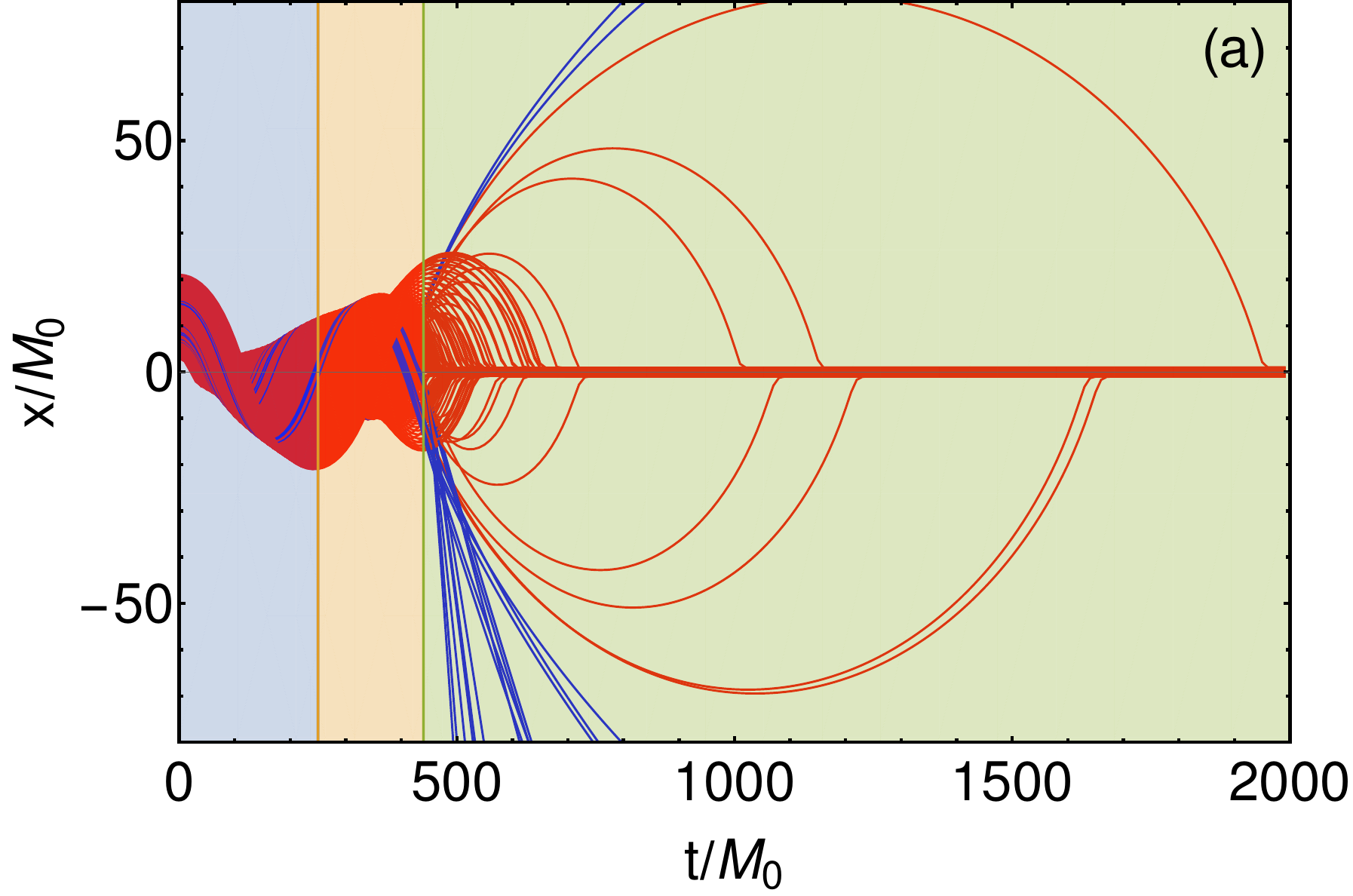} \hspace{5mm}
			\includegraphics[width=0.44\linewidth]{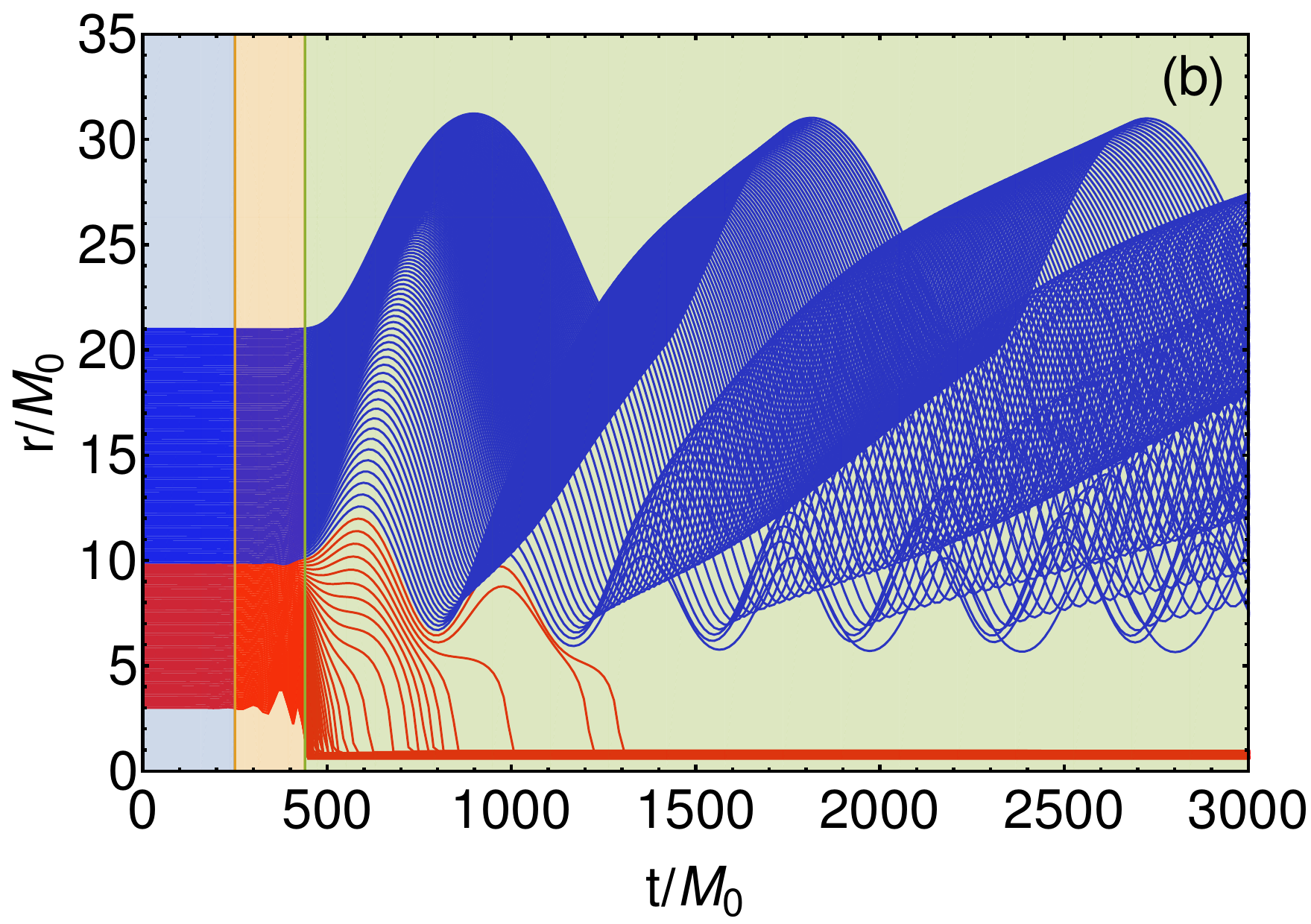} \hspace{5mm}
		\end{center}
		\vspace{-5mm}
		\caption{Evolution of the radial distance for the case BS\_b. The shaded regions are plotted in light blue for the quasi-equilibrium phase, in light orange for the oscillating and collapsing phase, and in light green for the black hole accretion growth phase. Blue lines denote geodesics that become unbound; red lines are geodesics that plunge into the black hole; and orange lines are for orbits that remain bounded but do not plunge. Panel (a) is for initially reciprocating orbits, and panel (b) is for initially circular orbits. Here, we choose $x=r \cos(\varphi)$.}
		\label{orb_088_l0}
	\end{figure*}

\begin{figure*}[htbp]
		\begin{center}
			\includegraphics[width=0.46\linewidth]{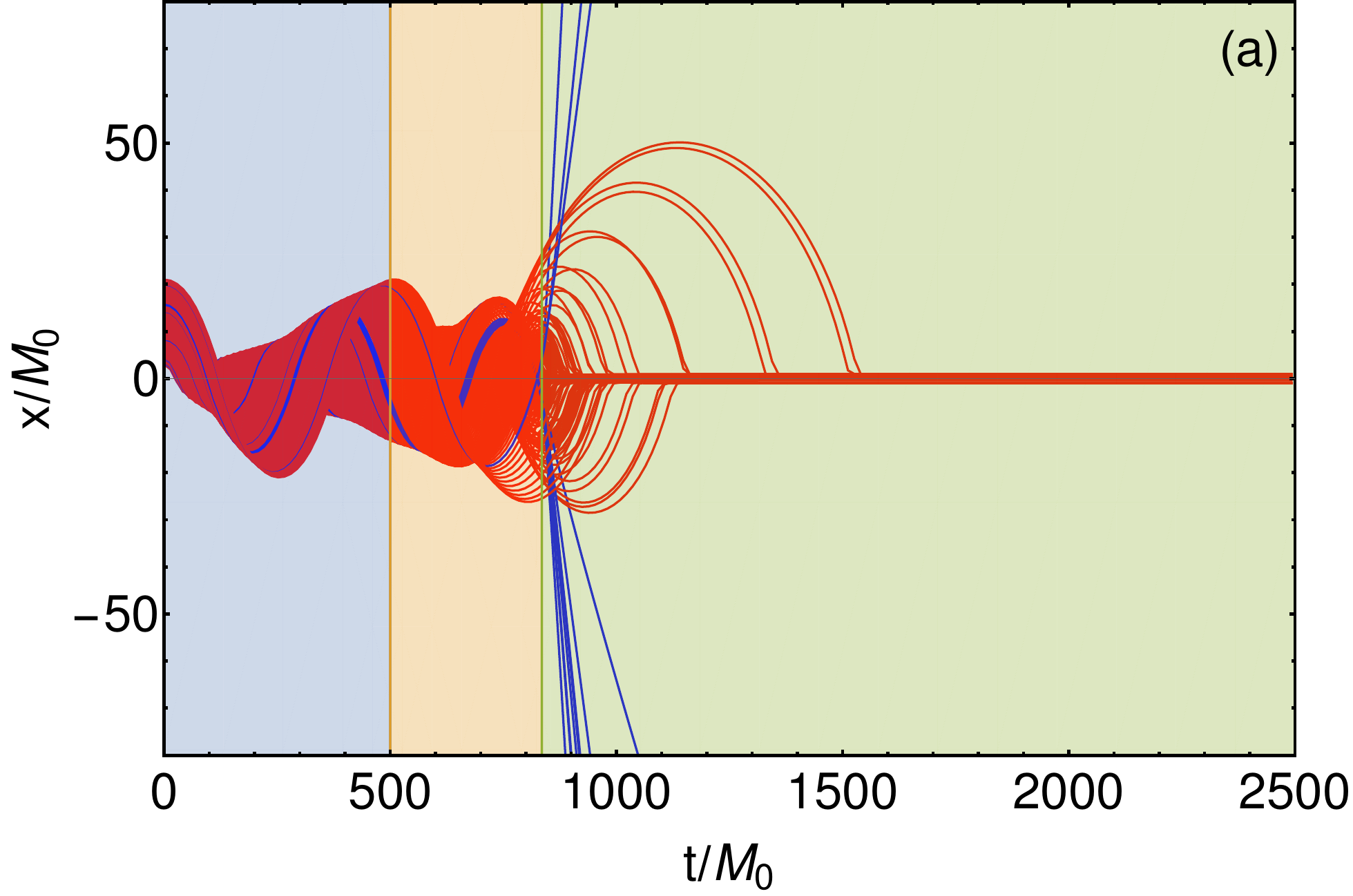}  \hspace{5mm}
			\includegraphics[width=0.45\linewidth]{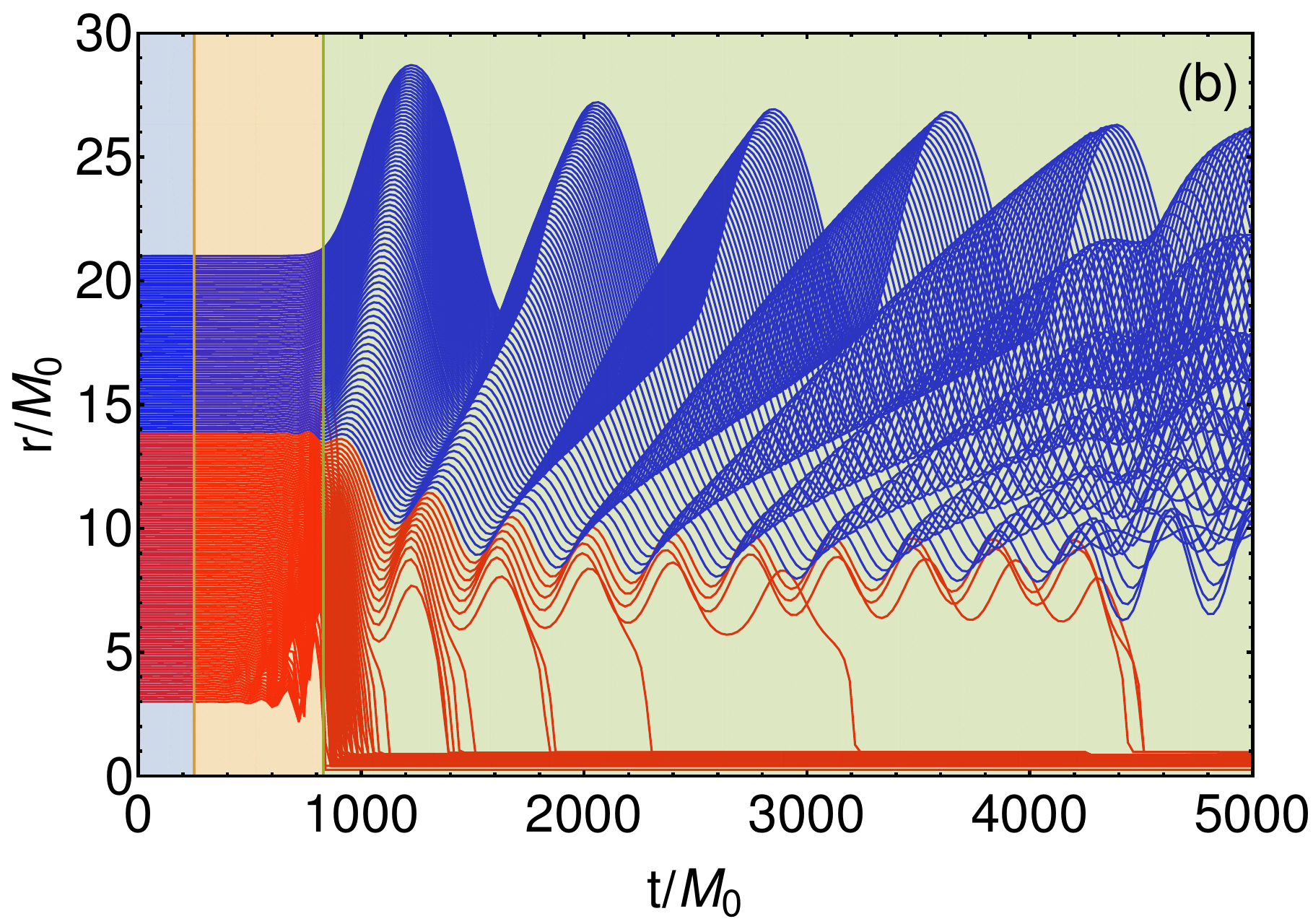} \hspace{5mm}
		\end{center}
		\vspace{-5mm}
		\caption{Same as Fig. \ref{orb_088_l0} but for the BS\_c case.}
		\label{orb_092_l0}
	\end{figure*}

The evolution of an unstable boson star collapsing into a black hole has three phases. The first is a quasi-equilibrium phase in which the space-time remains almost unchanged. In the second phase, the boson star oscillates and eventually collapses into a black hole. During the third phase, the black hole grows from accreating the remnant scalar field. Figures~\ref{orb_088_l0} and \ref{orb_092_l0} show the evolution of the radial coordinate of the geodesics for the cases BS\_b and BS\_c, respectively. The shaded regions are plotted as light blue for the quasi-equilibrium phase, light orange for the oscillating and collapsing phase, and light green for the black hole accretion growth phase. Blue lines denote geodesics that become unbound; red lines are geodesics that plunge into the black hole; and orange lines are for orbits that remain bounded but do not plunge. Panel (a) is for initially reciprocating orbits, and panel (b) is for initially circular orbits.

As seen in panel (a) of Figs.~\ref{orb_088_l0} and \ref{orb_092_l0}, reciprocating orbits remain reciprocating, that is, oscillating back and forth through the origin, during the quasi-equilibrium and collapsing phases. Once the black hole forms, some orbits plunge, and others become unbound depending on the corresponding initial maximal radii. {In the analysis of all reciprocating geodesic orbits, our findings indicate that it is impossible for bound orbits to persist around the newly formed black holes due to the absence of orbital angular momenta. All such orbits will either plunge into or escape from the newly formed black hole, a fate determined by their initial maximum radii. Notably, there exists no critical radius that distinguishes between orbits that plunge and those that escape.} Whether an orbit plunges or not depends on the radial distance the orbit has at the moment the black hole forms. The formation of the black hole changes the effective potential; thus, the radial position of the geodesic relative to the new location of the potential barrier determines the fate of the orbit, i.e., plunge or unbound.
	
For orbits initially circular, we observe from panel (b) in Figs.~\ref{orb_088_l0} and \ref{orb_092_l0} that they remain circular during the quasi-equilibrium and collapsing phases. {When they enter the black hole growth phase, the orbits will plunge if the initial radius satisfies $r/M_0 < 9.9$ for BS\_b and $r/M_0 < 13.8$ for BS\_c. When the initial radius satisfies $r/M_0 \ge 9.9$ for BS\_b and $r/M_0 \ge 13.9$ for BS\_c, the orbits will remain bound but become eccentric. Here we observe a critical radius for separating the finally unbound plunging orbits and the finally bound eccentric orbits. The outcomes here have similar reasons to the reciprocating case, namely the location of the potential barrier and height relative to the position of the geodesic at the time of black hole formation.}

In summary, the qualitative behaviors of the orbits for BS\_b and BS\_c are similar. The specific differences are due to the duration of the phases in each case.

\begin{figure*}[htbp]
		\begin{center}
			\includegraphics[width=0.46\linewidth]{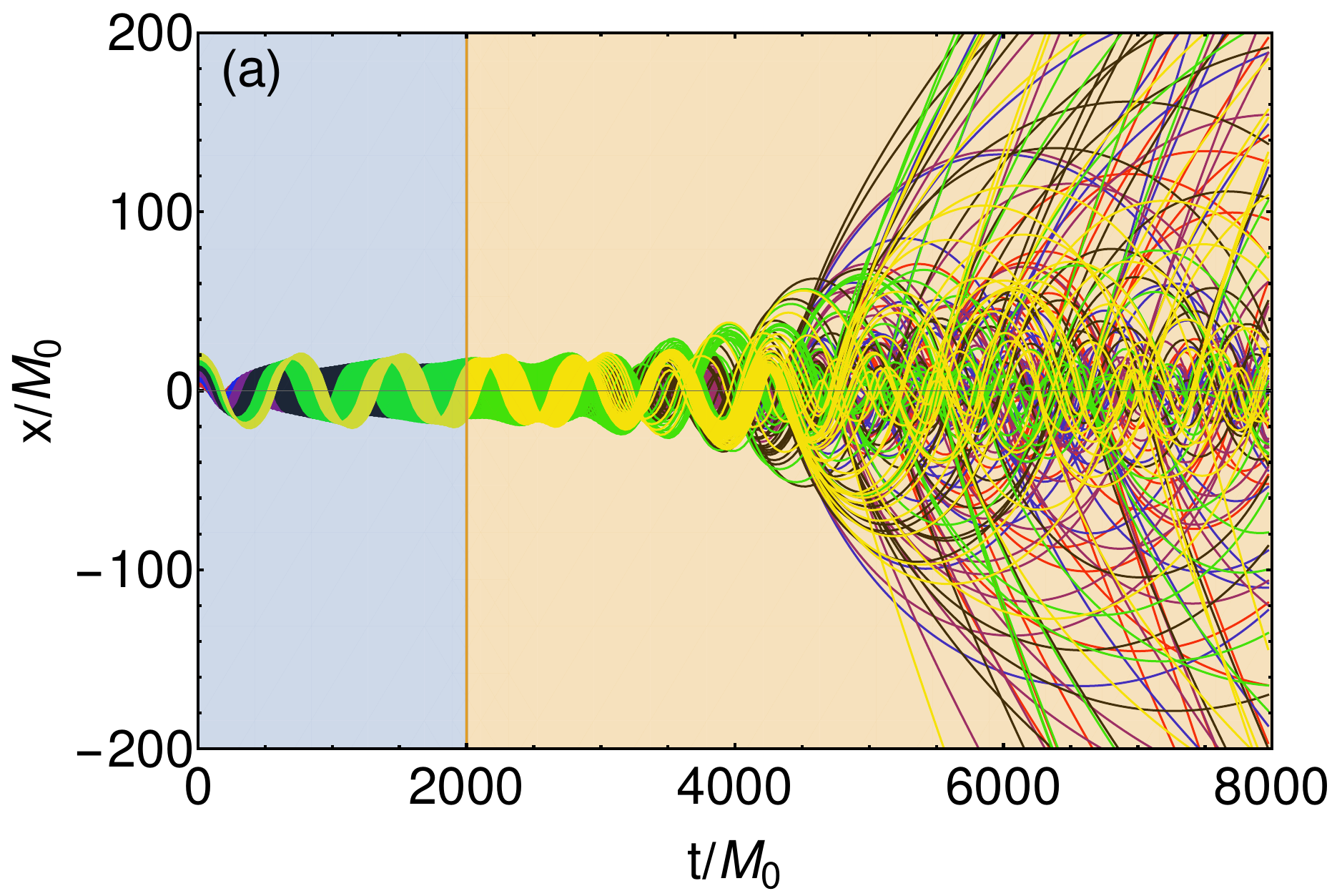} \hspace{5mm}
			\includegraphics[width=0.44\linewidth]{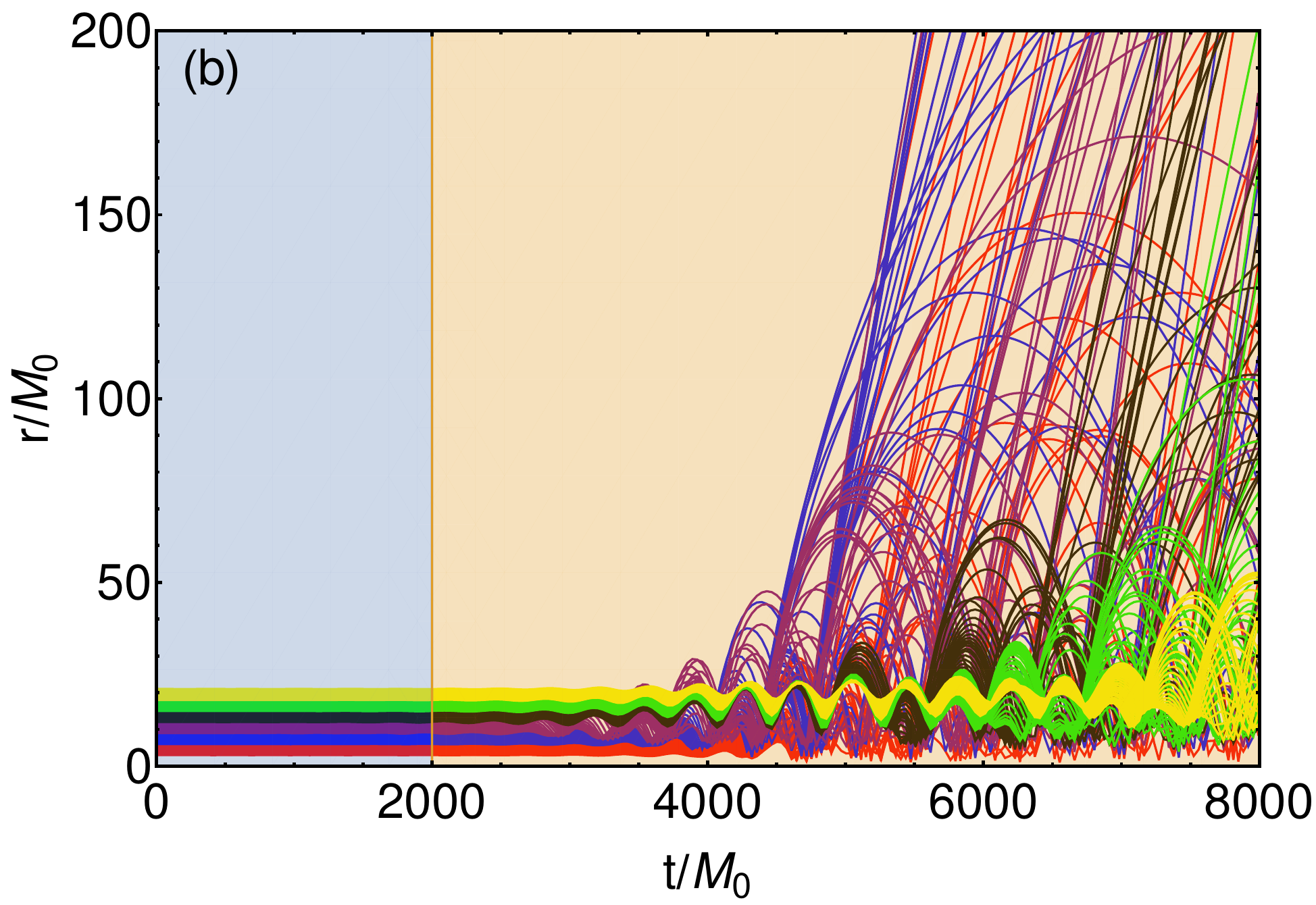} \hspace{5mm}
		\end{center}
		\vspace{-5mm}
		\caption{Same as Fig. \ref{orb_088_l0} but for the BS\_d case. Here the light blue background region is the quasi-equilibrium phase, the light orange background region is the oscillating and migrating phase. Here, we choose $x=r \cos(\varphi)$.}
		\label{orb_096_l50}
	\end{figure*}

Finally, we discuss the results for the unstable migrating case BS\_d, where the boson star migrates to a new equilibrium state and a black hole does not form. Figure \ref{orb_096_l50} shows the evolution of the geodesic radial distance for this case. In this situation there are two distinct boson star evolution phases. As the previous case, the first phase is also the quasi-equilibrium phase with the space-time almost unchanged. The second phase is the proper oscillating and migrating phase, with the space-time changing until it settles down into a new equilibrium state. {Nonetheless, ascertaining the very final state within a finite time is not feasible, as referenced in \cite{Sanchis-Gual:2021phr}. Consequently, we can only present the trajectories of the initially bound geodesics over an extensive yet finite duration. To vividly illustrate the ultimate fates of the initially bound reciprocating and circular orbits at later stages, we employ six distinct colors to differentiate the orbits based on their radii.} For all the reciprocating geodesic orbits, we do not observe a critical radius separating bound orbits from unbound ones. As seen in panel (a) of Fig.~\ref{orb_096_l50}, bound and unbound orbits occur at both small and large radii. For all the initial stable circular orbits, we observe  similar behaviors to that of the reciprocating orbits. Those that remain bound will have nozero eccentricities at late time. Such behaviors are completely different from the collapsing case.

\section{Conclusions}

{In this paper, we utilized numerical relativity combined with the $3+1$ formalism of geodesic equations to first present the fate of initially bound timelike geodesic orbits under spherical dynamical space-times of boson stars. We focused on initially bound circular and reciprocating orbits and investigated their final fate and dynamical behavior within the stable, unstable collapsing, and unstable migrating spherical boson stars. The geodesic integration was carried out simultaneously with the numerical evolution of the space-times.

In the stable boson star case, we verified that the state of the orbits remain unchanged. In the two unstable boson stars that collapse into black holes, {we discovered that initially reciprocating orbits and initially circular orbits exhibit distinct fates. For initially reciprocating orbits, all such orbits become unbound, with their ultimate fate either plunging or escaping that depending on their initial maximum radii. Notably, there is no critical radius to distinguish between orbits that plunge and those that escape. On the other hand, for initially circular orbits, those with smaller initial radii are observed to plunge into the newly formed black hole, while orbits with larger radii continue to orbit the newly formed black hole, displaying non-zero orbital eccentricities.} For the migrating boson star, there was no clear pattern; orbits either escape or remain bound but become eccentric. The study has given us a picture of the gravitational collapse and migration of boson stars in terms of the dynamics of geodesics. As expected, the changes in the space-times of the boson stars reflect on the changes in the effective potential that governs the dynamics of the geodesics.

It should be noted that we only investigated the dynamical changes of the initially bound timelike orbits within the unstable spherical boson stars. When considering the rotating boson stars or other multipole boson stars, the final states of the initially bound orbits in the above process will inevitably change. However, regardless of the symmetry of the boson stars, the unstable boson stars will always either collapse into black holes or migrate to other stable states. Therefore, compared to the spherical background, the final state in these cases may be accompanied by more details or nuances. The regions with separating, bound, unbound, and plunging orbits could potentially offer new insights into the behavior of matter in the vicinity of these objects, which could, in turn, be reflected in the emitted gravitational waves.

Apart from unstable boson stars, the phenomena such as black hole superradiance instability and its dynamical scalarization also depict the dynamic transition from hairless to hairy black hole, corresponding to changes in space-time during the transition of black holes from hairless to hairy states. Consequently, the properties of initially bound orbits in the background of initial hairless black holes are expected to undergo changes during the dynamical formation processes of hairy black holes. However, if we consider nonspherically symmetric systems, conducting the above-mentioned research would be technically extremely challenging. We will pursue corresponding investigations in our future work.}
	
\section{Acknowledgments}

We warmly thank Pedro V. P. Cunha for useful discussions and suggestions. This work was supported in part by the National Key Research and Development Program of China (Grant No. 2021YFC2203003), the National Natural Science Foundation of China (Grants No. 12105126, No. 12075103, and No 12247101), the 111 Project under (Grant No. B20063), the China Postdoctoral Science Foundation (Grant No. 2021M701531), the Major Science and Technology Projects of Gansu Province, and Lanzhou City's scientific research funding subsidy to Lanzhou University. P.L. was supported by NSF awards 2207780, 2114581, and 2114582.


\begin{thebibliography}{100}
		
		\bibitem{Abbott2016a}
		B. P. Abbott \textit{et al.} [LIGO Scientific Collaboration and Virgo Collaboration], {\em GW150914: The Advanced LIGO Detectors in the Era of First Discoveries}, Phys. Rev. Lett. \textbf{116}, 061102 (2016), [arXiv:1602.03838 [gr-qc]].
		
		\bibitem{LIGOScientific:2018jsj}
		B. P. Abbott \textit{et al.} [LIGO Scientific Collaboration and Virgo Collaboration],
		{\em Binary Black Hole Population Properties Inferred from the First and Second Observing Runs of Advanced LIGO and Advanced Virgo},
		Astrophys. J. Lett. \textbf{882}, L24 (2019), [arXiv:1811.12940 [astro-ph.HE]].
		
		\bibitem{LIGOScientific:2020kqk}
		R.~Abbott \textit{et al.} [LIGO Scientific Collaboration and Virgo Collaboration],
		{\em Population Properties of Compact Objects from the Second LIGO-Virgo Gravitational-Wave Transient Catalog},
		Astrophys. J. Lett. \textbf{913}, L7 (2021), [arXiv:2010.14533 [astro-ph.HE]].
		
		\bibitem{LIGOScientific:2021psn}
		R.~Abbott \textit{et al.} [LIGO Scientific, VIRGO and KAGRA],
		{\em The population of merging compact binaries inferred using gravitational waves through GWTC-3},
		Phys. Rev. X \textbf{13}, 011048 (2023), [arXiv:2111.03634 [astro-ph.HE]].
		
		\bibitem{eth2019}
		K. Akiyama \textit{et al.} [Event Horizon Telescope], {\em First M87 Event Horizon Telescope Results. I. The Shadow of the Supermassive Black Hole}, Astrophys. J. Lett. \textbf{875}, L1 (2019), [arXiv:1906.11238 [astro-ph.GA]]; {\em First M87 Event Horizon Telescope Results. II. Array and Instrumentation}, Astrophys. J. Lett. \textbf{875}, L2 (2019), [arXiv:1906.11239 [astro-ph.IM]]; {\em First M87 Event Horizon Telescope Results. III. Data Processing and Calibration}, Astrophys. J. Lett. \textbf{875}, L3 (2019); {\em First M87 Event Horizon Telescope Results. IV. Imaging the Central Supermassive Black Hole}, Astrophys. J. Lett. \textbf{875}, L4 (2019), [arXiv:1906.11241 [astro-ph.GA]]; {\em First M87 Event Horizon Telescope Results. V. Physical Origin of the Asymmetric Ring}, Astrophys. J. Lett. \textbf{875}, L5 (2019), [arXiv:1906.11242 [astro-ph.GA]]; {\em First M87 Event Horizon Telescope Results. VI. The Shadow and Mass of the Central Black Hole}, Astrophys. J. Lett. \textbf{875}, L6 (2019), [arXiv:1906.11243 [astro-ph.GA]].
		
		\bibitem{eth2022}
		K.~Akiyama \textit{et al.} [Event Horizon Telescope],
		{\em First Sagittarius A* Event Horizon Telescope Results. I. The Shadow of the Supermassive Black Hole in the Center of the Milky Way}, Astrophys. J. Lett. \textbf{930}, L12 (2022), [arXiv:2311.08680 [astro-ph.HE]]; {\em First Sagittarius A* Event Horizon Telescope Results. II. EHT and Multiwavelength Observations, Data Processing, and Calibration}, Astrophys. J. Lett. \textbf{930}, L13 (2022), [arXiv:2311.08679 [astro-ph.HE]]; {\em First Sagittarius A* Event Horizon Telescope Results. III. Imaging of the Galactic Center Supermassive Black Hole}, Astrophys. J. Lett. \textbf{930}, L14 (2022), [arXiv:2311.09479 [astro-ph.HE]]; {\em First Sagittarius A* Event Horizon Telescope Results. IV. Variability, Morphology, and Black Hole Mass}, Astrophys. J. Lett. \textbf{930}, L15 (2022), [arXiv:2311.08697 [astro-ph.HE]]; {\em First Sagittarius A* Event Horizon Telescope Results. V. Testing Astrophysical Models of the Galactic Center Black Hole}, Astrophys. J. Lett. \textbf{930}, L16 (2022), [arXiv:2311.09478 [astro-ph.HE]]; {\em First Sagittarius A* Event Horizon Telescope Results. VI. Testing the Black Hole Metric}, Astrophys. J. Lett. \textbf{930}, L17 (2022), [arXiv:2311.09484 [astro-ph.HE]].
	
		\bibitem{Feinblum1968}
		D. A. Feinblum and W. A. McKinley,
		{\em Stable states of a scalar particle in its own gravational field},
		Phys. Rev. \textbf{168}, 1445 (1968).
		
		\bibitem{Kaup1968}
		D. J. Kaup, {\em Klein-Gordon Geon},
		Phys. Rev. \textbf{172}, 1331 (1968).
		
		\bibitem{Ruffini1969}
		R. Ruffini and S. Bonazzola,
		{\em Systems of Self-Gravitating Particles in General Relativity and the Concept of an Equation of State},
		Phys. Rev. \textbf{187}, 1767 (1969).
		
		\bibitem{Seidel1994}
		E. Seidel and W.-M. Suen,
		{\em Formation of solitonic stars through gravitational cooling},
		Phys. Rev. Lett. \textbf{72}, 2516 (1994).
		
		\bibitem{Giovanni2018}
		F. D. Giovanni, N. Sanchis-Gual, C. A. R. Herdeiro, and J. A. Font,
		{\em Dynamical formation of Proca stars and quasistationary solitonic objects,}
		Phys. Rev. D \textbf{98}, 064044 (2018), [arXiv:1803.04802 [gr-qc]].
		
		\bibitem{Sanchis2019}
		N. Sanchis-Gual, F. D. Giovanni, M. Zilh\~ao, C. A. R. Herdeiro, P. Cerda-Duran, J. A. Font, and E. Radu,
		{\em Nonlinear Dynamics of Spinning Bosonic Stars: Formation and Stability},
		Phys. Rev. Lett. \textbf{123}, 221101 (2019), [arXiv:1907.12565 [gr-qc]].
		
		\bibitem{Cardoso:2019rvt}
		V.~Cardoso and P.~Pani,
		{\em Testing the nature of dark compact objects: a status report},
		Living Rev. Rel. \textbf{22}, 4 (2019), [arXiv:1904.05363 [gr-qc]].
		
		\bibitem{Bustillo2021}
		J. C. Bustillo, N. Sanchis-Gual, A. Torres-Forn\'e, J. A. Font, A. Vajpeyi, R. Smith, C. A. R. Herdeiro, E. Radu, and S. H. W. Leong,
		{\em GW190521 as a Merger of Proca Stars: A Potential New Vector Boson of $8.7\times10^{-13} eV$},	
		Phys. Rev. Lett. \textbf{126}, 081101 (2021), [arXiv:2009.05376 [gr-qc]].
		
		\bibitem{LIGOScientific:2020iuh}
		R.~Abbott \textit{et al.} [LIGO Scientific Collaboration and Virgo Collaboration],
		{\em GW190521: A Binary Black Hole Merger with a Total Mass of $150  M_{\odot}$},
		Phys. Rev. Lett. \textbf{125}, 101102 (2020), [arXiv:2009.01075 [gr-qc]].
		
		\bibitem{Siemonsen:2020hcg}
		N.~Siemonsen and W.~E.~East,
		{\em Stability of rotating scalar boson stars with nonlinear interactions},
		Phys. Rev. D \textbf{103}, 044022 (2021), [arXiv:2011.08247 [gr-qc]].
		
		\bibitem{Choptuik:2019zji}
		M.~Choptuik, R.~Masachs, and B.~Way,
		{\em Multioscillating Boson Stars},
		Phys. Rev. Lett. \textbf{123}, 131101 (2019), [arXiv:1904.02168 [gr-qc]].
		
		\bibitem{Sanchis-Gual:2021edp}
		N.~Sanchis-Gual, F.~D. Giovanni, C. A. R. Herdeiro, E.~Radu, and J.~A.~Font,
		{\em Multifield, Multifrequency Bosonic Stars and a Stabilization Mechanism},
		Phys. Rev. Lett. \textbf{126}, 241105 (2021), [arXiv:2103.12136 [gr-qc]].
		
		\bibitem{Sanchis-Gual:2021phr}
		N.~Sanchis-Gual, C. A. R. Herdeiro, and E.~Radu,
		{\em Self-interactions can stabilize excited boson stars},
		Class. Quant. Grav. \textbf{39}, 064001 (2022), [arXiv:2110.03000 [gr-qc]].
		
		\bibitem{Penrose1965}
		R. Penrose, {\em Gravitational collapse and space-time singularities}, Phys. Rev. Lett. \textbf{14}, 57 (1965).
		
		\bibitem{Wald1984}
		R. M. Wald, {\em General Relativity}, The University of Chicago Press (1984).
		
		\bibitem{1995Sci...270..941M}
		R. A. Matzner, H. E. Seidel, S. L. Shapiro, L. Smarr, W. M.~Suen, S. A. Teukolsky, and J. Winicour,
		{\em Geometry of a Black Hole Collision}",
		Science \textbf{270}, 941 (1995).
		
		\bibitem{Vincent:2012kn}
		F.~H.~Vincent, E.~Gourgoulhon, and J.~Novak,
		{\em $3+1$ geodesic equation and images in numerical spacetimes},
		Class. Quant. Grav. \textbf{29}, 245005 (2012), [arXiv:1208.3927 [gr-qc]].
		
		\bibitem{Bohn:2014xxa}
		A.~Bohn, W.~Throwe, F.~H\'ebert, K.~Henriksson, D.~Bunandar, M.~A.~Scheel, and N.~W.~Taylor,
		{\em What does a binary black hole merger look like?},
		Class. Quant. Grav. \textbf{32}, 065002 (2015), [arXiv:1410.7775 [gr-qc]].
		
		\bibitem{Herdeiro2021lwl}
		C.~A.~R.~Herdeiro, A.~M.~Pombo, E.~Radu, P.~V.~P.~Cunha, and N.~Sanchis-Gual,
		{\em The imitation game: Proca stars that can mimic the Schwarzschild shadow},
		JCAP \textbf{04}, 051 (2021), [arXiv:2102.01703 [gr-qc]].
		
		\bibitem{philippe2014}
		P. Grandcl\'ement, Claire Som\'e, and Eric Gourgoulhon,
		{\em Models of rotating boson stars and geodesics around them: New type of orbits},
		Phys. Rev. D, \textbf{90}, 024068, (2014), [arXiv:1405.4837 [gr-qc]].
		
		\bibitem{Grould12017}
		M. Grould, Z. Meliani, F. H. Vincent, P. Grandcl\'ement, and E. Gourgoulhon,
		{\em Comparing timelike geodesics around a Kerr black hole and a boson star},
		Class. Quan. Grav. \textbf{34}, 215007 (2017), [arXiv:1709.05938 [astro-ph.HE]].
		
		\bibitem{Collodel2018}
		L. G. Collodel, B. Kleihaus, and J.~A.~Kunz,
        {\em Static Orbits in Rotating Spacetimes},
		Phys. Rev. Lett. \textbf{120}, 201103 (2018), [arXiv:1711.05191 [gr-qc]].
		
		\bibitem{Yuzhang2021}
		Y.-P. Zhang, Y.-B. Zeng, Y.-Q. Wang, S.-W. Wei, and Y.-X. Liu,
		{\em Motion of test particle in rotating boson star},
		Phys. Rev. D \textbf{105}, 044021 (2022), [arXiv:2107.04848 [gr-qc]].
		
		\bibitem{Cunha:2017qtt}
		P.~V.~P.~Cunha, E.~Berti, and C.~A.~R.~Herdeiro,
		{\em Light-Ring Stability for Ultracompact Objects},
		Phys. Rev. Lett. \textbf{119}, 251102 (2017), [arXiv:1708.04211 [gr-qc]].
		
		\bibitem{Cunha:2022gde}
		P.~V.~P.~Cunha, C. A. R. Herdeiro, E.~Radu and N.~Sanchis-Gual,
		{\em Exotic Compact Objects and the Fate of the Light-Ring Instability},
		Phys. Rev. Lett. \textbf{130}, 061401 (2023), [arXiv:2207.13713 [gr-qc]].
		
		\bibitem{Montero:2012yr}
		P.~J.~Montero and I.~Cordero-Carrion,
		{\em BSSN equations in spherical coordinates without regularization: vacuum and non-vacuum spherically symmetric spacetimes},
		Phys. Rev. D \textbf{85}, 124037 (2012), [arXiv:1204.5377 [gr-qc]].
		
		\bibitem{Bona:1997hp}
		C.~Bona, J.~Masso, E.~Seidel, and J.~Stela,
		{\em First order hyperbolic formalism for numerical relativity},
		Phys. Rev. D \textbf{56}, 3405 (1997), [arXiv:gr-qc/9709016 [gr-qc]].
		
		\bibitem{Alcubierre:2002kk}
		M.~Alcubierre, B.~Bruegmann, P.~Diener, M.~Koppitz, D.~Pollney, E.~Seidel, and R.~Takahashi,
		{\em Gauge conditions for long term numerical black hole evolutions without excision},
		Phys. Rev. D \textbf{67}, 084023 (2003), [arXiv:gr-qc/0206072 [gr-qc]].
		
		\bibitem{Kreiss-Oliger}
		H.-O. Kreiss and J. Oliger,
        {\em Methods for the Approximate Solution of the Time Dependent Problems}
        (GARP, Geneva, 1973).
		
	\end{thebibliography}
\end{document}